\documentclass[english,english,english]{aa}
\usepackage[T1]{fontenc}
\usepackage[latin1]{inputenc}
\usepackage{babel}
\usepackage{graphicx}

\usepackage{times}
\usepackage[T1]{fontenc}
\usepackage[latin1]{inputenc}
\usepackage{babel}
\usepackage{graphics}
\usepackage{txfonts}

\usepackage{natbib} 
\bibpunct{(}{)}{;}{a}{}{,} % to follow the A&A style 

\begin{document}

\title{The diffuse nature of Str\"{o}mgren spheres}

\author{J. Ritzerveld}

\offprints{J. Ritzerveld}

\institute{Sterrewacht Leiden, P.O. Box 9513, 2300 RA Leiden, The Netherlands
\\
 \email{ritzerveld@strw.leidenuniv.nl}}

\date{Received ; Accepted}

\abstract{In this Letter, we argue that the standard analytical derivations of properties of HII regions, such as the speed, shape and asymptotic position of ionisation fronts require a more precise treatment. These derivations use the \emph{on the spot approximation}, which in effect ignores the diffuse component of the radiation field. We show that, in fact, HII regions are diffusion dominated. This has as a result that the morphology of inhomogeneous HII regions will be drastically different, because shadowing effects have a less profound impact on the apparent shape. Moreover, it will have influence on the propagation speed of ionisation fronts. We quantify our claims by analytically deriving the internal radiation structure of  HII regions, taking diffusion fully into account for several different cosmologically relevant density distributions.
}

\maketitle

\section{Introduction}

Ionized hydrogen is produced when the UV radiation emitted copiously by hot young stars ionizes the surrounding interstellar and intergalactic medium. 
Recently, the resultant HII regions have gained a lot of interest, because when the first hot and massive Pop III stars form, the emitted energetic photons blow ionisation bubbles, which after overlapping will eventually fill the Universe at the end of the Epoch of Reionisation. Therefore, much effort has been put into the physical, analytical and numerical understanding of the mechanisms involved.

A very relevant property of HII regions, is the structure of the diffuse radiation field, as this will influence the dynamics of the expansion, but also the morphology of shadows formed within the region. It is well known \citep{Hong, Lopez} that diffuse photons dominate near the Str\"{o}mgren radius, but for practical reasons this is often ignored, under the assumption that it is a small effect. In this Letter, we quantitatively show that it is a non-negligible effect, especially in the presence of density gradients.

\section{The On-The-Spot Approximation}

Most analytical work done on HII regions heavily depends on the \emph{on-the-spot approximation}, part of which has its origin in \cite{Baker}, and has been discussed and used in the standard texts \cite{Spitzer} and \cite{Osterbrock}. In this argument, one considers the gas within the ionisation front, or within the Str\"{o}mgren sphere when the static equilibrium solution has already been reached, and does some bookkeeping to equate the number of recombinations within that volume to the number of ionising photons. Electron captures directly into level \(n=1\), parametrised by the coefficient \(\alpha_\mathrm{1}(T)\), produce photons energetic enough to ionise another atom (the spectrum of these diffuse photons can be approximated by a delta function just above the Lyman limit). The ionisation balance can be drawn up as follows:

\begin{eqnarray}
\label{Stromgren}
  4\pi r^2 n(r) \frac{dr}{dt}&=&S_*+4\pi\alpha_\mathrm{1}(T)\int n^2(r)r^2dr \nonumber\\
  & - & 4\pi\alpha_\mathrm{A}(T)\int n^2(r)r^2dr 
\end{eqnarray}
in which \(r\) is the position of the front and \(S_*\) is the number of ionising photons (i.e. with frequencies \(\nu\ge\nu_0\) above the Lyman limit treshold) emitted by the source per second. The second term on the right-hand side is the total number of diffuse photons, given the HI density distribution \(n(r)\). The last term on the right-hand side accounts for the total number of recombinations to every level, parametrised by the recombination coefficient \(\alpha_\mathrm{A}(T)\). The integrations are over the whole HII region. If the Str\"{o}mgren radius has already been reached, the three contributions on the rhs of Eqn.(\ref{Stromgren}) cancel, but up until that point the ionisation front has a finite speed \(dr/dt\).

Unlike the source photons, which are directed outwards, the diffuse photons may cross the nebula in any direction from their point of creation. Thus, one has to do a full treatment of radiative transfer, which complicates matters drastically. This is where the on-the-spot approximation enters the picture. For typical HII regions, the physical parameters are such that the optical depth for photons near the Lyman limit is \(\sim 30\) \cite[cf.][]{Osterbrock}. Thus, one argues that the diffuse photons produced by recombinations to the ground level will be re-absorbed very close to where they were produced (on the spot). This has as a result that effectively the recombinations directly to the ground level do not count, because they are exactly balanced by the photons they produce. Henceforth, one can ignore these recombinations, and the diffuse photons created by them, and use the effective \emph{Case B} \citep[using terminology introduced by][]{Baker} recombination coefficient

\begin{equation}
\label{caseB}
  \alpha_\mathrm{B}(T)=\alpha_\mathrm{A}(T)-\alpha_\mathrm{1}(T),
\end{equation}
with Eqn.(\ref{Stromgren}) reducing to

\begin{equation}
\label{StromgrenB}
  4\pi r^2 n(r) \frac{dr}{dt}=S_*+4\pi\alpha_\mathrm{B}(T)\int n^2(r)r^2dr.
\end{equation}
The Str\"{o}mgren sphere radius can be easily obtained from Eqn.(\ref{StromgrenB}) by considering the equilibrium solution with zero velocity of the ionisation front \(dr/dt=0\).

We will argue that it is not permitted to use the on the spot approximation for anything more than just calculating the radius of the Str\"{o}mgren sphere. First and foremost, the on the spot approximation is based on the argument that the mean free path for Lyman limit photons is very small, and that thus all diffuse photons are locally re-absorbed. But this argument is not correct, because the mean free path for the source photons (assuming these are also near the Lyman limit) is just as small. Thus, the diffuse and the source photons are on equal footing energy-wise, and no distinction between them can be made, excluding their directionality. Thus, at each location \(r\) within the HII region, not only are the diffuse photons used to compensate for the \(\alpha_\mathrm{1}\) recombinations, but also a weighted fraction of the source photons, by which a fraction of the diffuse photons can actually escape. In effect, a fraction of the mono-directional source photons is converted into diffuse radiation.

One can therefore see that the number of directional source photons decreases, when moving outwards from the source, until at a certain point the diffuse radiation starts to dominate. In the next section, we will analytically derive the diffuse versus mono-directional structure of the radiation within a Str\"{o}mgren sphere for several astrophysically important density distributions.

\section {The diffuse structure within Str\"{o}mgren spheres}
In this section, we will derive analytically the diffuse and source radiation structure within a spherically symmetric HII region which is already in equilibrium, i.e. the Str\"{o}mgren radius has already been reached. In subsequent sections, we will use these results to draw conclusions for more general cases.

\subsection{Homogeneous HI matter distribution}
In the following, we assume a homogeneous hydrogen medium density \(n_\mathrm{HI}(r)=n_\mathrm{HI}\), a temperature \(T=10^4\mathrm{K}\) for the ionised plasma, and, following the considerations that led up to the on the spot approximation, we assume that the mean free path of the photons is very small, until locally the gas is fully ionised. We do the calculations for a Str\"{o}mgren sphere in \(d\)-dimensional space, because we can use the general result in the following subsections.

Because the Str\"{o}mgren radius has already been reached, we know that there is an equilibrium between the number of recombinations per timestep and the total number of ionising photons emitted by the source and the gas, so we may drop the time-dependence. The number of recombinations to \emph{every} level and the number of diffuse photons produced within a spherical shell at distance \(r\) is

\begin{equation}
\label{NumRecomb}
  \left\{\begin{array}{ccl}
   N_\mathrm{rec}(r)&=&\xi(d)\alpha_\mathrm{A}n^2r^{d-1}dr \\
   N_\mathrm{diff}(r)&=&\xi(d)\alpha_\mathrm{1}n^2r^{d-1}dr,  
   \end{array}\right.
\end{equation}
in which \(\xi=2\pi^{1/2}/\Gamma(d/2)\) (\(=4\pi\) in 3D) is just a geometrical factor and \(\Gamma(x)\) is the Gamma function. The number of photons emitted by the source plus the number of diffuse photons within the whole region must compensate all the recombinations, from which we obtain

\begin{eqnarray}
\label{SourceNum}
  S_*&=&\xi(d)\alpha_\mathrm{B}n^2\int^{R_\mathrm{S}}_0 r^{d-1}dr \nonumber\\
  &=&\frac{\xi(d)}{d}\alpha_\mathrm{B}n^2R^d_\mathrm{S},
\end{eqnarray}
in which \(R_\mathrm{S}\) is the Str\"{o}mgren radius.

If we define \(I_\mathrm{s}(r)\) and \(I_\mathrm{d}(r)\) as the number of source and diffuse photons per unit volume, respectively, left at radius \(r\le R_\mathrm{S}\), we can write down the following coupled system of differential equations on the domain \(0\le r\le R_\mathrm{S}\)

\begin{equation}
 \label{System}
  \left\{ \begin{array}{ccl}
            \frac{dI_\mathrm{s}(r)}{dr}&=&-\xi(d)\alpha_\mathrm{A}n^2r^{d-1}{I_\mathrm{s}(r)}\left({I_\mathrm{s}(r)+I_\mathrm{d}(r)}\right)^{-1} \\
            \frac{dI_\mathrm{d}(r)}{dr}&=&-\xi(d)\alpha_\mathrm{A}n^2r^{d-1}{I_\mathrm{d}(r)}\left({I_\mathrm{s}(r)+I_\mathrm{d}(r)}\right)^{-1}+\xi(d)\alpha_\mathrm{1}n^2r^{d-1}
            \end{array}\right..
\end{equation}
The first term on the rhs of both equations is the total number of recombinations being compensated for via a weighing term, that takes care of the fact that, if there are more source than diffuse photons present, the recombinations will mainly be compensated by the source photons, and vice versa. This system of equations is closed via the initial conditions \(I_\mathrm{d}(0)=0\) and \(I_\mathrm{s}(0)=S_*\). We should note that we have used a symmetry condition here, in the sense that the diffuse radiation sent into the opposite direction (away from the front) is exactly balanced by the diffuse radiation on the other side of the source.

If we add the two equations, the weighing terms add up to unity, and using the initial conditions, we easily obtain an equation for the total radiation field

\begin{equation}
  \label{Total}
  (I_\mathrm{s}+I_\mathrm{d})(r)=\frac{\xi(d)}{d}\alpha_\mathrm{B}n^2R^d_\mathrm{S}\left(1-\left(\frac{r}{R_\mathrm{S}}\right)^d\right).
\end{equation}
We can plug this into the weighing terms in Eqs.(\ref{System}), and upon integrating we obtain

\begin{equation}
  \label{Source_Frac}
  I_\mathrm{s}(r)=(I_\mathrm{s}+I_\mathrm{d})(r)-I_\mathrm{d}(r)=\frac{\xi(d)}{d}\alpha_\mathrm{B}n^2R^d_\mathrm{S}\left(1-\left(\frac{r}{R_\mathrm{S}}\right)^d\right)^{{\alpha_\mathrm{A}}/{\alpha_\mathrm{B}}}.
\end{equation}

\begin{figure}
{\begin{center}\includegraphics[width=6cm]{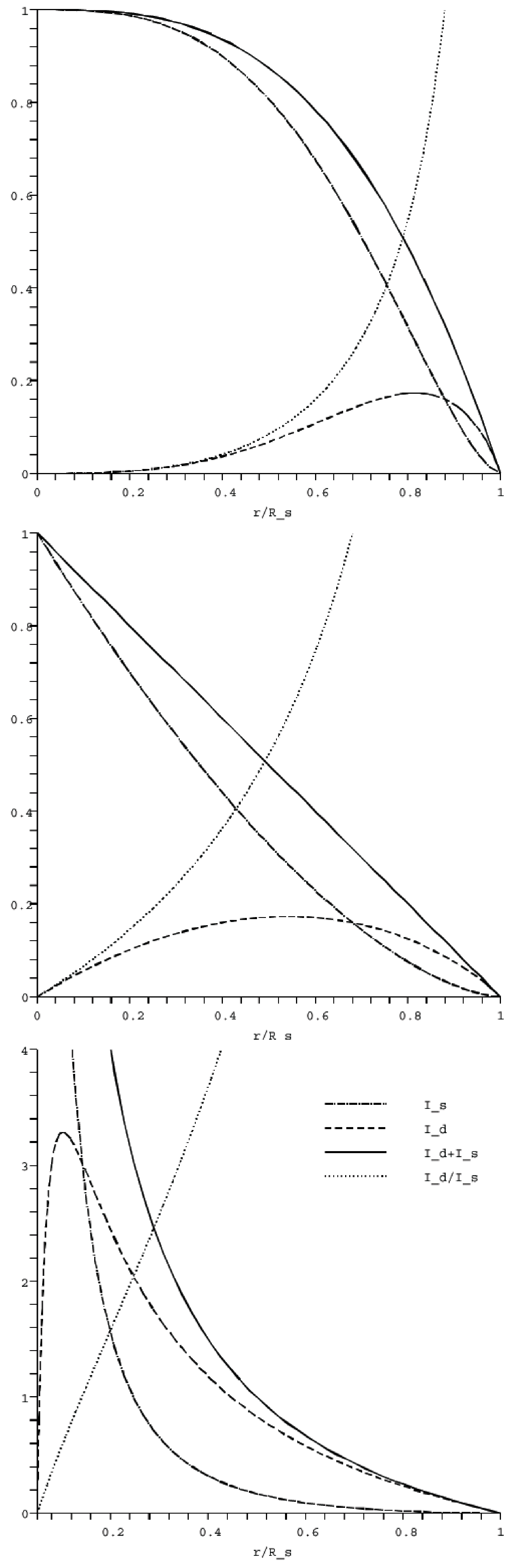}\end{center}}
\caption{\label{GraphsFig}Graphs of the functions \((I_\mathrm{s}+I_\mathrm{d})(r)\) (\emph{solid}), \(I_\mathrm{s}\) (\emph{dashed-dotted}), \(I_\mathrm{d}\) (\emph{dashed}), and \(I_\mathrm{d}/I_\mathrm{s}\) (\emph{dotted}) for a homogeneous (\emph{top}), an \(r^{-1}\) (\emph{centre}), and an \(r^{-2}\) (\emph{bottom}) medium distribution.}
\end{figure}
After choosing units such that \(\frac{\xi(d)}{d}\alpha_\mathrm{B}n^2R^d_\mathrm{S}=1\), and using the ratio \(\alpha_\mathrm{A}/\alpha_\mathrm{B}=4.18/2.60\) at \(10^4\mathrm{K}\) \cite[e.g.][]{Osterbrock}, we can draw a figure that plots \((I_\mathrm{s}+I_\mathrm{d})(r)\), \(I_\mathrm{s}(r)\), \(I_\mathrm{d}(r)\) and the ratio \(I_\mathrm{d}(r)/I_\mathrm{s}(r)\). This has been done for the normal \(d=3\) space in Fig. \ref{GraphsFig}, top. One can see that at a certain radius \(r_\mathrm{eq}\) the diffuse radiation component takes over. We can derive \(r_\mathrm{eq}\) exactly by solving \(I_\mathrm{s}(r)=I_\mathrm{d}(r)\), and obtain

\begin{equation}
  \label{Equiv}
  r_\mathrm{eq}=R_\mathrm{S}\left(1-2^{-\alpha_\mathrm{B}/\alpha_\mathrm{1}}\right)^{1/d}.
\end{equation}

Filling in the values results for 3D in \(r_\mathrm{eq}/R_\mathrm{S}=87.6\%\). Thus, we can immediately conclude that for a homogeneous gas distribution the outer \(12\%\) of the HII region is dominated by diffuse radiation, and that very near the front only diffuse photons are present.  This means that, in terms of volume, \(33\%\) of the HII region is diffusion dominated. This effect is even greater for higher temperatures, for which the ratio \(\alpha_\mathrm{B}/\alpha_\mathrm{1}\) is smaller.

\subsection{An \(r^{-1}\) HI gas distribution}
The effect described in the previous section is even more drastic for a HI density distribution in the form \(n(r)=n_c(r/r_c)^{-1}\). This density profile has been used to describe the central regions of a dark matter halo \citep{Navarro}.

Following the same arguments as in the previous Subsection, we get for the 3D case
\begin{equation}
\label{NumRecombInv}
  \left\{\begin{array}{lcl}
   N_\mathrm{rec}(r)&=&4\pi\alpha_\mathrm{A}n^2_cr^2_cdr \\
   N_\mathrm{diff}(r)&=&4\pi\alpha_\mathrm{1}n^2_cr^2_cdr \\
   S_*&=& 4\pi\alpha_\mathrm{B}n^2_cr^2_cR_\mathrm{S},
   \end{array}\right.
\end{equation}
which are similar to Eqs.(\ref{NumRecomb}) and (\ref{SourceNum}) for \(d=1\), but with a different constant factor. Thus, we can use Eqs.(\ref{Total}), (\ref{Source_Frac}) and (\ref{Equiv}), and obtain
\begin{equation}
\label{InvSol}
  \left\{\begin{array}{lcl}
 (I_\mathrm{s}+I_\mathrm{d})(r)&=&4\pi\alpha_\mathrm{B}n^2_cr^2_cR_\mathrm{S}\left(1-\frac{r}{R_\mathrm{S}}\right) \\
 I_\mathrm{s}(r)&=&4\pi\alpha_\mathrm{B}n^2_cr^2_cR_\mathrm{S}\left(1-\frac{r}{R_\mathrm{S}}\right)^{\frac{\alpha_\mathrm{A}}{\alpha_\mathrm{B}}} \\
r_\mathrm{eq}&=&R_\mathrm{S}\left(1-2^{-\alpha_\mathrm{B}/\alpha_\mathrm{1}}\right).
   \end{array}\right.
\end{equation}

The graphs of these functions are plotted in the centre part of Fig. \ref{GraphsFig}. Filling in the value for \(\alpha_\mathrm{B}/\alpha_\mathrm{1}\) at \(T=10^4\mathrm{K}\), gives \(r_\mathrm{eq}/R_\mathrm{S}=67.2\%\), from which we can conclude that for an isotropic \(r^{-1}\) distributed HII region the outer \(33\%\) is diffusion dominated, which is equivalent to roughly \(70\%\) of the total volume. 

\subsection{An \(r^{-2}\) HI gas distribution}
The effect is strongest for a HI density distribution in the form \(n(r)=n_c(r/r_c)^{-2}\), which has, for example, been used to describe the density profile of halos just interior to the virial radius and of collapsing isothermal spheres. In the derivation for this density distribution, we have to be more careful, because the equivalents of Eqs.(\ref{NumRecomb}) and (\ref{SourceNum}) will contain singularities at \(r=0\), which can be resolved by choosing an inner edge \(0\le r_0 \le R_\mathrm{S}\). Thus, we obtain

\begin{equation}
\label{NumRecombInvSq}
  \left\{\begin{array}{lcl}
   N_\mathrm{rec}(r)&=&4\pi\alpha_\mathrm{A}n^2_cr^4_cr^{-2}dr \\
   N_\mathrm{diff}(r)&=&4\pi\alpha_\mathrm{1}n^2_cr^4_cr^{-2}dr \\
   S_*&=& 4\pi\alpha_\mathrm{B}n^2_cr^4_c\left(r^{-1}_0 - R^{-1}_\mathrm{S}\right).
   \end{array}\right.
\end{equation}
Using similar steps as in Subsection 3.1, we obtain
\begin{equation}
\label{InvSolSq}
  \left\{\begin{array}{lcl}
 (I_\mathrm{s}+I_\mathrm{d})(r)&=&4\pi\alpha_\mathrm{B}n^2_cr^4_c \left(r^{-1}-R^{-1}_\mathrm{S}\right) \\
 I_\mathrm{s}(r)&=&4\pi\alpha_\mathrm{B}n^2_cr^4_c\left(r^{-1}_0-R^{-1}_\mathrm{S}\right)^{1-\frac{\alpha_\mathrm{A}}{\alpha_\mathrm{B}}}\left(r^{-1}-R^{-1}_\mathrm{S}\right)^{\frac{\alpha_\mathrm{A}}{\alpha_\mathrm{B}}} \\
r_\mathrm{eq}&=&R_\mathrm{S}\left(1+2^{-\alpha_\mathrm{B}/\alpha_\mathrm{1}}\left(R_\mathrm{S}/r_0-1\right)\right)^{-1}.
   \end{array}\right.
\end{equation}
The graphs of these functions are plotted in the bottom part of Fig. \ref{GraphsFig} for \(r_0\le r\le R_\mathrm{S}\). Taking a typical value of \(r_0/R_\mathrm{S}=0.05\), we find that \(r_\mathrm{eq}/R_\mathrm{S}=13.8\%\), which means that, in terms of volume, \(99.7\%\) of the HII region is diffusion dominated.

\subsection{Hard photons}
\begin{figure}
{\begin{center}\includegraphics[width=7cm]{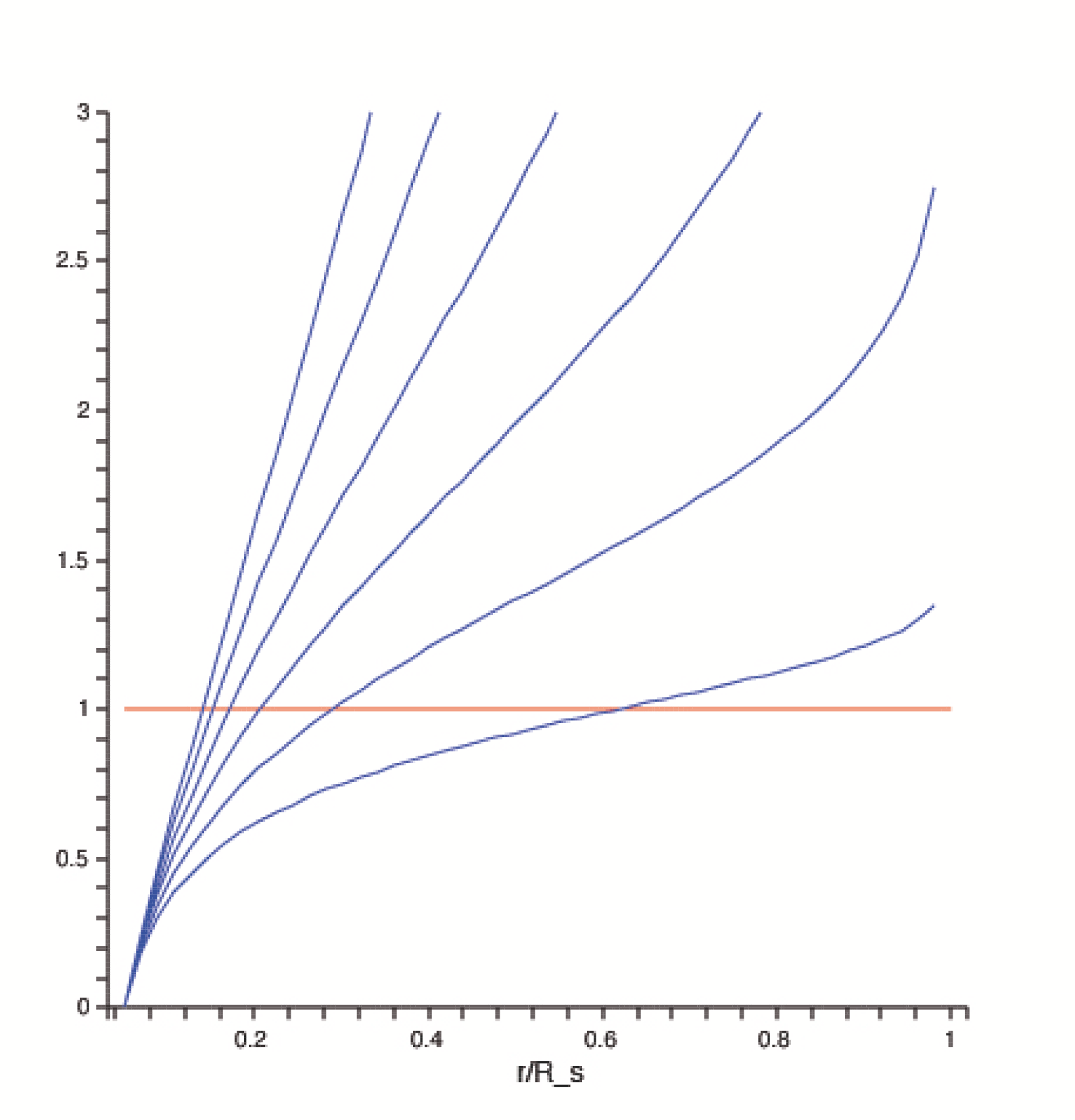}\end{center}}
\caption{\label{GraphsCFig}Graphs of the functions \((I_\mathrm{d}/I_\mathrm{s})(r)\) for the varying values (from top to bottom) \(c=1.0,0.9,0.8,0.7,0.6,0.5\) in a \(r^{-2}\) density distribution. Overplotted is the line \(I_\mathrm{d}/I_\mathrm{s}=1\) which determines the radius above which the diffusion dominates.}
\end{figure}

The analysis in the previous subsections was done assuming that the source spectrum is peaked just above the Lyman limit. It is much more realistic to assume that a substantial part of the luminosity will be in the higher frequency domain. These photons have a longer mean free path (\(a(\nu)\propto\nu^{-3}\)) than the diffuse photons, from which we expect that their contribution will in effect move \(r_\mathrm{eq}\)  towards \(R_\mathrm{S}\). To accurately model this, we would have to do a full radiative transfer treatment, but we can give an analytical estimate of this effect. Hereto, we do not take the source spectrum to be a delta function just above the Lyman limit, but at a certain higher frequency. As a result, we relax the condition that the medium is optically very thick to source photons, by which not \(I_\mathrm{s}(r)\) photons contribute to the locally available radiation field, but only a factor \(0\le c \le1\) thereof. Nothing changes in the equations except that the weighing factors will now be \(cI_\mathrm{s}/(cI_\mathrm{s}+I_\mathrm{d})\) and \(I_\mathrm{d}/(cI_\mathrm{s}+I_\mathrm{d})\). We solved the equations for the \(r^{-2}\) distribution for varying values of \(c\) and plotted the results in Fig. \ref{GraphsCFig}. One can see that the position of \(r_\mathrm{eq}\) does change, but that the diffuse radiation field effect is still very noticeable.

\section{Concluding}
What does all of this analysis entail? First and foremost, it shows that a large part of HII regions, in some cases even almost the whole volume, is diffusion dominated. This has a profound effect on the formation of shadows behind clumps within the HII region. A more quantitative analysis on this matter has to be done in forthcoming work, but one can immediately see that, even in the case of the only moderately diffuse homogenous medium distribution (subsection 3.1), if a dense clump is put within the radius \(r_\mathrm{eq}\), the resulting shadow will be squeezed by the diffuse field in the outer region. This effect will be much stronger than in the analysis in \cite{Canto}, in which only the locally produced diffuse photons can move into the shadow region, whereas in our analysis also the overall diffuse radiation field will have a profound influence. This effect will have an impact on (cosmological) radiative transfer codes which use ray tracing methods to delineate sharp shadows, and the resultant \(21\mathrm{cm}\) signatures of the HI regions in between the overlapping (re)ionisation bubbles, which may be observed in the near future.

Another effect is that the diffuse radiation field will influence the dynamics of the HII region. From the considerations in the previous sections for the equilibrium solution, one can see that the only radiation component pushing against the front is the diffuse one, which is not mono-directional, but isotropic. Thus, the net flux ionising photons in the direction of the front is less. The symmetry condition, as used in subsection 3.1, is only valid in the equilibrium (time-\emph{in}dependent) solution, but in the time-dependent case the speed of photons diffusing from one side of the HII region to the other is finite, and, as a result, this effect may slow down the ionisation front, by which the standard analytical solutions to Eqn.(\ref{Stromgren}), as found in e.g. \cite{Spitzer}, will have to be modified to take into account that this diffusion speed is considerably smaller than the speed of light. If it turns out that this effect can not be neglected, it will have an influence on the analysis of how fast the first stars and/or quasars could have reionised the Universe.

Concluding, the usual on-the-spot approximation in which the diffuse radiation component is neglected can only be used to calculate the Str\"{o}mgren sphere within an isotropic HI medium distribution. We have shown that the diffuse component can not be neglected, which will result in a drastic change of the morphology of anistropic distributions, and which may change the speed of the ionisation fronts.

\begin{acknowledgements}
I would like to thank Vincent Icke and Garrelt Mellema for carefully reading this manuscript; I also thank Harm Habing for enlightening discussions, and finally Sijme-Jan Paardekoper for his mathematical insight. I also thank the referee W. Steffen for valuable comments. This work was carried out with support from grant 635.000.009 from the Netherlands Organisation for Scientific Research.
\end{acknowledgements}

% for the bibliography, at the end 
\bibliographystyle{aa} % style aa.bst 
\bibliography{Ritzerveld} 

\end{document}